\documentclass[superscriptaddress,amsmath,amssymb,aps,prb,twocolumn]{revtex4-2}
\usepackage{graphicx}         
\usepackage{dcolumn}          
\usepackage{bm}               
\usepackage{upgreek}
\usepackage{booktabs} 				
\usepackage{tabularx}
\usepackage{array}
\usepackage{siunitx}
\usepackage[colorlinks=true,urlcolor=blue,breaklinks=true]{hyperref}
\usepackage{xcolor}
\usepackage{xr}

\begin{document}

\title{Supercurrent in Bi$_4$Te$_3$ Topological Material-Based Three-Terminal Junctions}

\author{Jonas K\"{o}lzer}
\affiliation{Peter Gr\"unberg Institut (PGI-9), Forschungszentrum J\"ulich, Wilhelm-Johnen-Straße, 52425 J\"ulich, Germany}
\affiliation{JARA-Fundamentals of Future Information Technology, J\"ulich-Aachen Research Alliance, Forschungszentrum J\"ulich and RWTH Aachen University, 52425 J\"ulich, Germany}

\author{Abdur Rehman Jalil}
\affiliation{Peter Gr\"unberg Institut (PGI-9), Forschungszentrum J\"ulich, Wilhelm-Johnen-Straße, 52425 J\"ulich, Germany}
\affiliation{JARA-Fundamentals of Future Information Technology, J\"ulich-Aachen Research Alliance, Forschungszentrum J\"ulich and RWTH Aachen University, 52425 J\"ulich, Germany}

\author{Daniel Rosenbach}
\affiliation{Peter Gr\"unberg Institut (PGI-9), Forschungszentrum J\"ulich, Wilhelm-Johnen-Straße, 52425 J\"ulich, Germany}
\affiliation{JARA-Fundamentals of Future Information Technology, J\"ulich-Aachen Research Alliance, Forschungszentrum J\"ulich and RWTH Aachen University, 52425 J\"ulich, Germany}

\author{Lisa Arndt}
\affiliation{JARA Institute for Quantum Information, RWTH Aachen University, Germany}

\author{Gregor Mussler}
\affiliation{Peter Gr\"unberg Institut (PGI-9), Forschungszentrum J\"ulich, Wilhelm-Johnen-Straße, 52425 J\"ulich, Germany}
\affiliation{JARA-Fundamentals of Future Information Technology, J\"ulich-Aachen Research Alliance, Forschungszentrum J\"ulich and RWTH Aachen University, 52425 J\"ulich, Germany}

\author{Peter Sch\"uffelgen}
\affiliation{Peter Gr\"unberg Institut (PGI-9), Forschungszentrum J\"ulich, Wilhelm-Johnen-Straße, 52425 J\"ulich, Germany}
\affiliation{JARA-Fundamentals of Future Information Technology, J\"ulich-Aachen Research Alliance, Forschungszentrum J\"ulich and RWTH Aachen University, 52425 J\"ulich, Germany}

\author{Detlev Gr\"utzmacher}
\affiliation{Peter Gr\"unberg Institut (PGI-9), Forschungszentrum J\"ulich, Wilhelm-Johnen-Straße, 52425 J\"ulich, Germany}
\affiliation{JARA-Fundamentals of Future Information Technology, J\"ulich-Aachen Research Alliance, Forschungszentrum J\"ulich and RWTH Aachen University, 52425 J\"ulich, Germany}

\author{Hans L\"uth}
\affiliation{Peter Gr\"unberg Institut (PGI-9), Forschungszentrum J\"ulich, Wilhelm-Johnen-Straße, 52425 J\"ulich, Germany}
\affiliation{JARA-Fundamentals of Future Information Technology, J\"ulich-Aachen Research Alliance, Forschungszentrum J\"ulich and RWTH Aachen University, 52425 J\"ulich, Germany}

\author{Thomas Sch\"apers}
\email{th.schaepers@fz-juelich.de}
\affiliation{Peter Gr\"unberg Institut (PGI-9), Forschungszentrum J\"ulich, Wilhelm-Johnen-Straße, 52425 J\"ulich, Germany}
\affiliation{JARA-Fundamentals of Future Information Technology, J\"ulich-Aachen Research Alliance, Forschungszentrum J\"ulich and RWTH Aachen University, 52425 J\"ulich, Germany}

\hyphenation{}
\date{\today}

\begin{abstract}
In an in-situ prepared  three-terminal Josephson junction based on the topological insulator Bi$_4$Te$_3$ and the superconductor Nb the transport properties are studied. The differential resistance maps as a function of two bias currents reveal extended areas of Josephson supercurrent including coupling effects between adjacent superconducting electrodes. The observed dynamics for the coupling of the junctions is interpreted using a numerical simulation of a similar geometry based on a resistively and capacitively shunted Josephson junction model. The temperature dependency indicates that the device behaves similar to prior experiments with single Josephson junctions comprising topological insulators weak links. Irradiating radio frequencies to the junction we find a spectrum of integer Shapiro steps and an additional fractional step, which is interpreted by a skewed current-phase relationship. In a perpendicular magnetic field we observe Fraunhofer-like interference patterns of the switching currents. 
\end{abstract}

\maketitle

\section{Introduction} \label{sec:introduction}

Hybrid structures comprising three-dimensional topological insulator nanoribbons combined with superconductors are a very promising platform for realizing circuits for fault-tolerant topological quantum computing \cite{Kitaev03,Hyart13,Aasen16,Manousakis17}. For its operation Majorana bound states are employed, which are formed by aligning an external magnetic field with a nanoribbon proximitized with an s-type superconductor \cite{Cook11,Cook12,Legg21}. For the braiding of different pairs of Majorana states for qubit operation multi-terminal structures are required \cite{Fu08,Hyart13,Stenger19}. Braiding can be performed by adjusting the superconducting phase of the superconducting electrodes to each other.   

Multi-terminal Josephson junctions are the backbone of Majorana braiding mechanism in a topological qubit; where a three-terminal Josephson junction acts as a basic building block \cite{Hyart13}. Understanding the superconducting transport in such a device holds a key importance for the realization of a topological quantum system. Generally, the use of hybrid devices with multiple connections leads to rich physics in terms of transport properties. Indeed, theoretical studies have investigated singularities, such as Weyl nodes, in the Andreev spectra of multi-terminal Josephson junctions \cite{Yokoyama15,Eriksson17,Xie18}. Moreover, multi-terminal Josephson junctions with topologically trivial superconducting leads may lead to realizations where the junction itself can be regarded as an artificial topological material \cite{Riwar16}. Furthermore, three-terminal junctions also allow transport via the quartet mechanism and non-local Andreev processes \cite{Houzet10,Freyn11,Nowak19,Melo22}. 

On the experimental side, multi-terminal Josephson junctions were fabricated with different materials for the weak link. In three-terminal Josephson junctions with a Cu or InAs nanowire subgap states \cite{Pfeffer14,Cohen18} and half-integer Shapiro steps \cite{Duvauchelle16} were observed, indicating transport via quartets of entangled Cooper pairs. Supercurrent flow affected by dissipative currents in an adjacent junction was studied on graphene-based junctions \cite{Draelos19}. Moreover, the higher-dimensional phase space was found to lead to fractional Shapiro steps in this type of junctions due to the inverse AC Josephson effect \cite{Arnault21}. By combining a multi-terminal junction with a top gate, the effect of gate voltage and magnetic field on the critical current contour has been studied \cite{Graziano20,Pankratova20,Graziano22}. Recently, flakes of the topological insulator Bi$_2$Se$_3$ were also used as a weak link in an interferometer structure, and evidence for a non-sinusoidal current-phase relationship was observed \cite{Kurter15}. In flux-controlled three-terminal junctions based on Bi$_2$Te$_3$, the opening and closing of a minigap was studied using normal probes \cite{Yang19}.

Here, we report on the transport properties of a three-terminal Josephson junction based on the Bi$_4$Te$_3$ material system as the weak link and Nb as the superconductor. To fabricate the samples, we used selective-area growth for the Bi$_4$Te$_3$ layer in combination with an in-situ bridge technology to define the superconducting electrodes \cite{Schueffelgen19}. Bi$_4$Te$_3$ is a natural superlattice of alternating Bi$_2$ bilayers and Bi$_2$Te$_3$ quintuple layers. Initially, Bi$_4$Te$_3$ has been reported to be a semimetal with zero band gap and a Dirac cone at the $\Gamma$ point \cite{Saito17}. However, recent band structure calculations in conjunction with scanning tunneling spectroscopy and angular photoemission spectroscopy measurements suggest that the material is a semimetal with topological surface states \cite{Chagas20,Chagas22,Nabok22}. In particular, advanced $GW$-band structure calculations have shown that a band gap of about $\SI{0.2}{\eV}$ opens at the $\Gamma$ point, which significantly reduces the density of the bulk state in this energy range \cite{Nabok22}. Bi$_4$Te$_3$ is classified as a dual topological insulator, a strong topological insulator with a non-zero mirror Chern number, i.e. a topological crystalline insulator phase. Though Bi$_4$Te$_3$ does not exhibit the proposed Dirac semimetal phase, it is still a very interesting material as it resides in close proximity to the critical point of band crossing in the topological phase diagram of Bi$_x$Te$_y$ alloys \cite{Jalil22}. Such a transition is proposed by Yang et al. \cite{Yang14a} where a topological crystalline insulator (Bi$_2$Te$_3$) \cite{Rauch14} can be topologically transformed into a topological Dirac semimetal through alloying it with other materials. On our multi-terminal junctions, we first investigated the DC properties and related the results to simulations based on the resistively and capacitively shunted Josephson junction (RCSJ) model. We then measured the radio frequency (rf) response, finding evidence for coupling of adjacent junctions. Finally, the behavior of our three-terminal junctions when an out-of-plane magnetic field is applied is investigated.  

\section{Experimental}

\begin{figure*}
  \centering
  \includegraphics[width=0.7\linewidth]{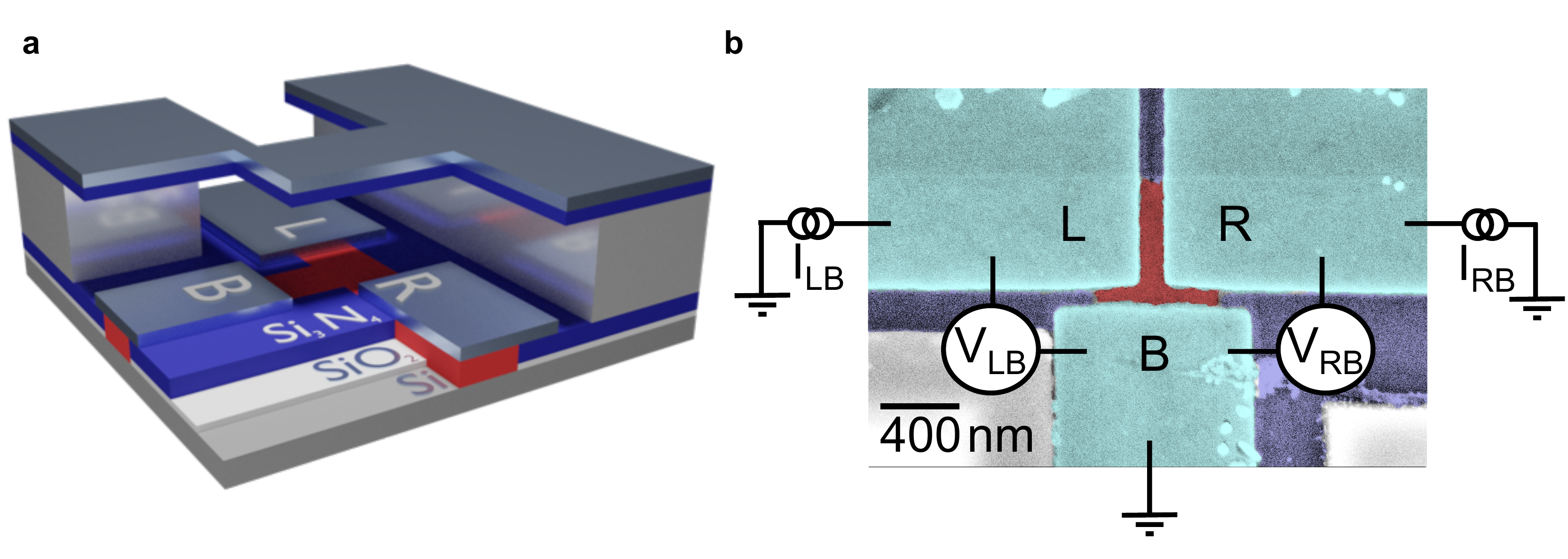}
  \caption{Rendering of a selective-area grown three-terminal Josephson junction and false color scanning electron micrograph with circuit: $(a)$ The three-terminal junction is composed of the silicon substrate (gray bottom layer), the first hard mask composed out of a silicon oxide (white)/ silicon nitride (blue) layer (as indicated by the labels). On top of this another hard mask layer composed of silicon oxide (white) and silicon nitride (blue) is deposited and patterned as a shadow mask. The topological insulator (red) is grown selectively into the first hard mask trench and the shadow mask is used for the definition of the junction in the metal deposition (silver) step. $(b)$ False-color scanning electron micrograph of the in-situ prepared three-terminal junction device. Niobium contacts (cyan) are deposited on top of the TI (red). The measurement configuration is also indicated.
  }
  \label{fig:Scheme_SEM}
\end{figure*}

Using the previously introduced technologies of topological insulator selective-area growth and in-situ bridge technology we fabricated three-terminal Josephson junctions, as illustrated in Fig.~\ref{fig:Scheme_SEM}(a) \cite{Rosenbach21,Schueffelgen19}. The geometry of the nanoribbon T-shaped junction for selective-area growth is defined by trenches in a SiO$_2$/Si$_3$N$_4$ (5\,nm/15\,nm) layer on a highly-resistive Si (111) substrate \cite{Schmitt22}. First, the 600-nm-wide nanotrenches are etched into the top Si$_3$N$_4$ layer using a combination of electron beam lithography and reactive ion etching. Subsequently, a second set of layers, i.e. a 100-nm-thick SiO$_2$ layer and a 300-nm-thick Si$_3$N$_4$ layer, is deposited on top to define the stencil mask for the in-situ Nb deposition \cite{Schueffelgen19}. After patterning the structures for the stencil mask into Si$_3$N$_4$, SiO$_2$ is etched in hydrofluoric acid (HF) forming the free-hanging bridge structures. Simultaneously, the Si(111) surface in the selective-area growth trenches is released in the bottom SiO$_2$ layer defined by the Si$_3$N$_4$ layer on top. The Bi$_4$Te$_3$ layer is selectively grown within these trenches, while the Si$_3$N$_4$ bridge structures are employed to define the geometry of the in situ deposited  superconducting electrodes \cite{Schueffelgen19}. The Bi$_4$Te$_3$ layer is grown  at a temperature of  $310^\circ$C using molecular beam epitaxy. Subsequently, the 50-nm-thick superconducting Nb electrodes are deposited by electron beam evaporation followed by covering the whole structure with a 5-nm-thick Al$_2$O$_3$ dielectric capping layer. Our processing scheme ensured a high-quality crystalline topological insulator material with clean superconductor interfaces~\cite{Koelzer21,Schueffelgen19}, as reported in previous transmission electron microscopy studies. An electron microscopy image of the investigated device is presented in Fig.~\ref{fig:Scheme_SEM}(b). 

The measurements of the three-terminal Josephson junction were carried out in a dilution refrigerator with base temperature of $T = 25\,\textnormal{mK}$. containing a $1\text{ - }1\text{ - }6\,\textnormal{T}$ vector magnet. As indicated in Fig.~\ref{fig:Scheme_SEM}(b), the left, right, and bottom junction electrodes are labeled as "L", "R", and "B", respectively. Two current sources supply currents $I_\mathrm{LB}$ and $I_\mathrm{RB}$ from L and R to the bottom electrode, respectively, with the according voltages $V_\mathrm{LB}$ and $V_\mathrm{RB}$ measured. The differential resistances are measured by adding an ac current of 10\,nA to the DC current bias using a lock-in amplifier. The rf-irradiation for the Shapiro step measurements was provided by an antenna placed in close vicinity to the sample.

\section{Results and Discussion} \label{sec:results}

\subsection*{DC characteristics}

\begin{figure}
  \centering
  \includegraphics[width=\linewidth]{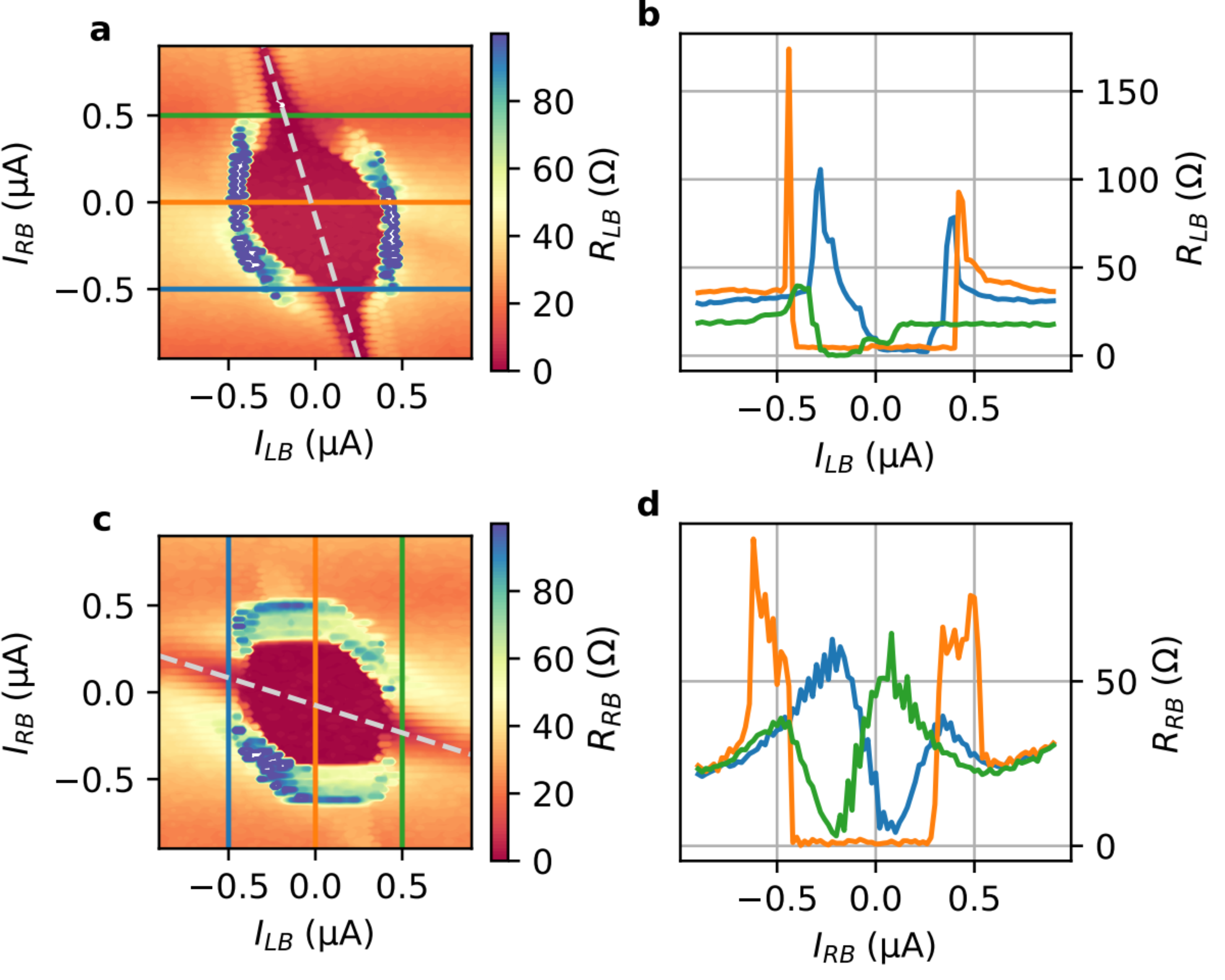}
  \caption{Differential resistance maps:
  (\textbf{a}) shows $R_\mathrm{LB}$ as a function of the bias currents $I_\mathrm{LB}$ and $I_\mathrm{RB}$ at 25\,mK with corresponding line cuts given in (\textbf{b}). In (\textbf{c}) the differential resistance map of $R_\mathrm{RB}$ is depicted with a selection of line cuts given in (\textbf{d}). The dashed lines in \textbf{a} and (\textbf{c}) indicate the superconducting regions of compensating bias currents. The differential resistances was measured by using lock-in technique, i.e.  $R_\mathrm{LB}= \Delta V_\mathrm{LB}/ \Delta I_\mathrm{LB}$ and $R_\mathrm{RB}=\Delta V_\mathrm{RB}/ \Delta I_\mathrm{RB}$. 
  }
  \label{fig:T_JJ_Rdiff}
\end{figure}

Information about the basic junction characteristics is obtained by measuring the differential resistances $R_\mathrm{LB}=\Delta V_\mathrm{LB}/ \Delta I_\mathrm{LB}$ 
and $R_\mathrm{RB}=\Delta V_\mathrm{RB}/ \Delta I_\mathrm{RB}$ as a function of the bias currents $I_\mathrm{LB}$ and $I_\mathrm{RB}$, respectively. Starting with the left junction we find that $R_\mathrm{LB}$ shown in Figs~\ref{fig:T_JJ_Rdiff}(a) and (b) contains a superconducting region in the center when $I_\mathrm{LB}$ and $I_\mathrm{RB}$ are varied. The observed critical current contour is similar to what has been observed in induced superconducting nano junctions made of high mobility materials such as InAs/Al~\cite{Graziano20,Pankratova20} or  graphene~\cite{Draelos19}. 
The superconducting region extends along an inclined line indicated by the dashed line in Fig.~\ref{fig:T_JJ_Rdiff}(a). The switching to the superconducting state can be seen in the line cuts at fix values $I_\mathrm{RB}=0$ and $\pm 0.7\mu$A provided in Fig.~\ref{fig:T_JJ_Rdiff}(b). The extension of the superconducting state originates from a part of $I_\mathrm{RB}$ which flows via R to L through the junction between L and B compensating the current $I_\mathrm{LR}$ partly and by that reducing the total current. For our three-terminal device no reduced differential resistance is observed along the line $I_\mathrm{LB}=I_\mathrm{RB}$, which would indicate the presence of a Josephson supercurrent between the junction formed between electrodes L and R \cite{Graziano20,Graziano22}. We attribute this to the fact that the distance between these electrodes is slightly larger than for the other junctions so that no Josephson supercurrent is obtained. However, the junction between L and R acts as a shunt resistor taking care that the switching to the superconducting state is non-hysteretic. The differential resistance $R_\mathrm{RB}$ measured between R and B electrodes, depicted in Figs~\ref{fig:T_JJ_Rdiff}(c) and (d), shows a similar behaviour as $R_\mathrm{LB}$, i.e. featuring also an extended superconducting range due a compensation provided by part of $I_\mathrm{LR}$. The tilt of the superconducting range indicated by the dashed line in Figure~\ref{fig:T_JJ_Rdiff}(c) is lower compared to Fig.~\ref{fig:T_JJ_Rdiff}(a) since now $I_\mathrm{LR}$ is the compensating current.      

\subsection*{Simulations}

\begin{figure}
  \centering
  \includegraphics[width=\linewidth]{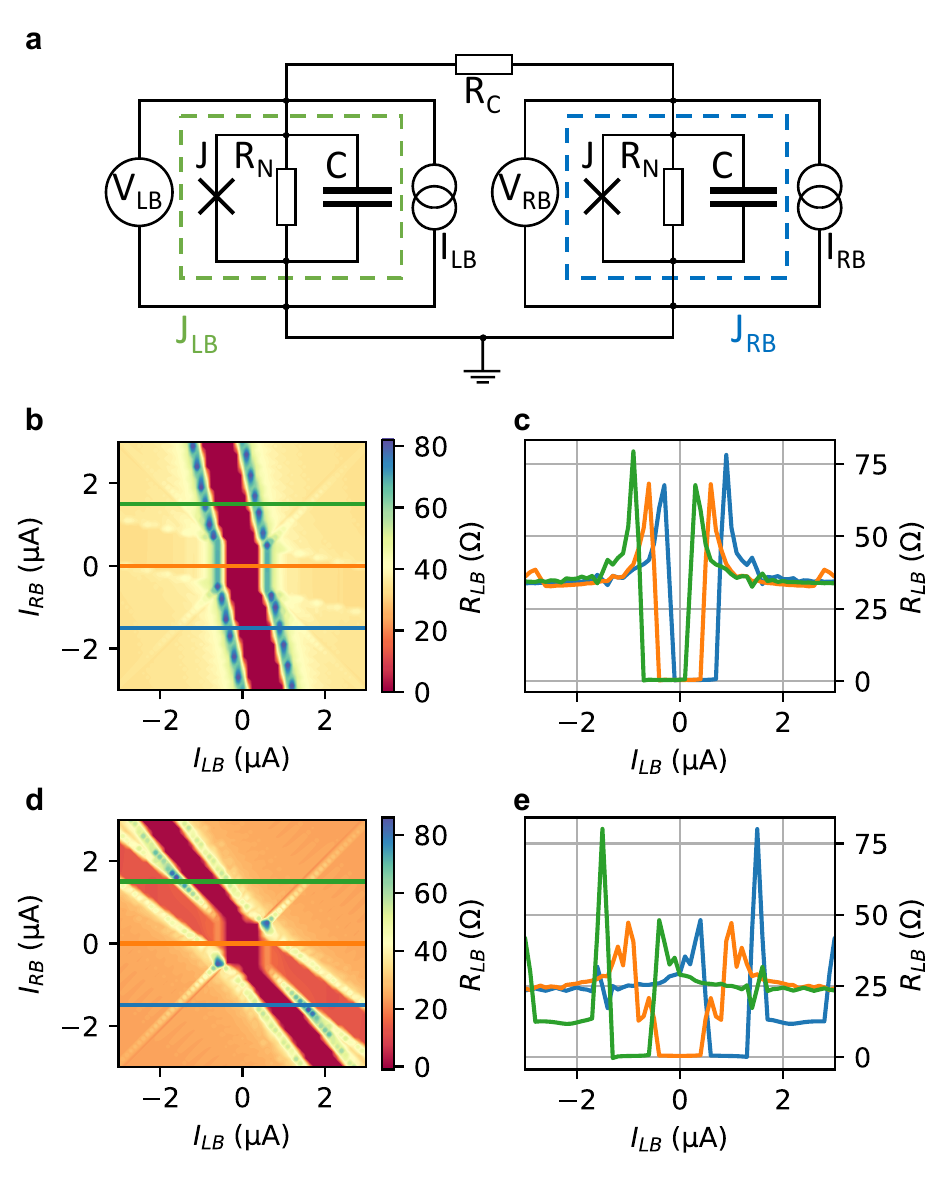}
  \caption{Numerical simulation of different coupling scenarios:
  (a) The three-terminal circuit is modeled by two RCSJ shunted Josephson junctions $J_\mathrm{LB}$ and $J_\mathrm{RB}$ (green and blue dashed line boxes), which are each modeled by a resistor $R_N$, a capacitor c, and an ideal Josephson junction $J$. Currents $I_\mathrm{LB}$ and $I_\mathrm{RB}$ are supplied via current sources while the voltage drops $V_\mathrm{LB}$ and $V_\mathrm{RB}$ across the junctions are measured. Both junctions are coupled via a coupling resistance $R_C$. (b) Differential resistance $R_\mathrm{LB}$ as a function of current biases for a realistic scenario for $R_C$ close to the one extracted in the experiment: $R_N=\SI{40}{\ohm}$, $R_C=\SI{160}{\ohm}$, $I_c=\SI{538}{nA}$, $\beta_c= (2e/\hbar) I_c R_N^2 C=0.1$. The zero resistance range is observed as a tilted line due to a compensation by a part of $I_\mathrm{RB}$.  Additionally, the influence of the second junction is observed as a similar line close to horizontal orientation.  The corresponding line cuts indicated in (b) are presented in (c). The scenario for a very small coupling resistance ($R_C \rightarrow 0$) is shown as a color map of $R_\mathrm{LB}$ and selected line cuts in (d) and (e).}
  \label{fig:T_JJ_sim_rc}
\end{figure}

The experimental results are modeled by assuming a network of two resistively and capacitively shunted Josephson (RCSJ) junctions coupled by a resistor $R_C$, as illustrated in Fig.~\ref{fig:T_JJ_sim_rc}(a). Solving the related system of differential equations numerically, in analogy to what was presented in previous works \cite{Graziano20,Arnault21}, we simulate the behaviour of the experimental system (information about the procedure see Supplementary Material). The results of the simulations are shown in Figs.~\ref{fig:T_JJ_sim_rc}(b) to (e), where the differential resistance $R_\mathrm{LB}$ is given as a function of the bias currents $I_\mathrm{LB}$ and $I_\mathrm{RB}$. 

The model describes the experiment well by reproducing the Josephson supercurrent along the inclined lines originating from compensating currents from both electrodes with a superconducting region at the center. The inclination is determined by the coupling resistance $R_C$. In Figs.~\ref{fig:T_JJ_sim_rc}(b) and (c), the coupling resistance was taken as $R_C=4 \cdot R_\mathrm{LB}$, with $R_\mathrm{LB}=\SI{40}{\ohm}$ which results in the same tilt as observed experimentally. Taking these values into account the normal state resistance is given by $R_N = 6/5 \cdot R_\mathrm{LB} = \SI{48}{\ohm}$. In our simulations for the critical current and for the Steward-McCumber parameter we assumed $I_c=\SI{538}{nA}$ and $\beta_c= (2e/\hbar) I_c R_N^2 C=0.1$, respectively, with c the junction capacitance. We found that the superconducting state in the junction between R and B leads to some weak feature as a similar line inclined towards horizontal orientation. Note, that for this line $R_\mathrm{LB}$ is non-zero, as the supercurrent in the other junction only partly reduces the current in the junction between L and B and hence only partially reduces the voltage drop. A noticeable difference between experiment and simulation is that in the measurements the extension of the superconducting state observed along the inclined line (cf. Fig.~\ref{fig:T_JJ_Rdiff}(a) is decreased compared to the simulation depicted in Fig.~\ref{fig:T_JJ_sim_rc}(b). As discussed by Draelos \textit{et al.} \cite{Draelos19}, this effect can be explained by dissipation in the neighboring junction being in the normal state resulting in an effective heating, in particular for junctions with small dimensions. In our simulation the direct coupling between the different junctions was neglected. As shown by Arnault \textit{et al.} \cite{Arnault21}, including coupling results in a more complex contour of the critical current area. If the coupling resistance becomes very small, i.e. $R_C \rightarrow 0$, the observed lines in the differential resistance shift towards the diagonal (cf. Figs.~\ref{fig:T_JJ_sim_rc}(d) and (e). Thus, both junctions are maximally correlated to both current biases $I_\mathrm{LB}$ and $I_\mathrm{RB}$.

\subsection*{Temperature dependence}
\begin{figure}
  \centering
  \includegraphics[width=\linewidth]{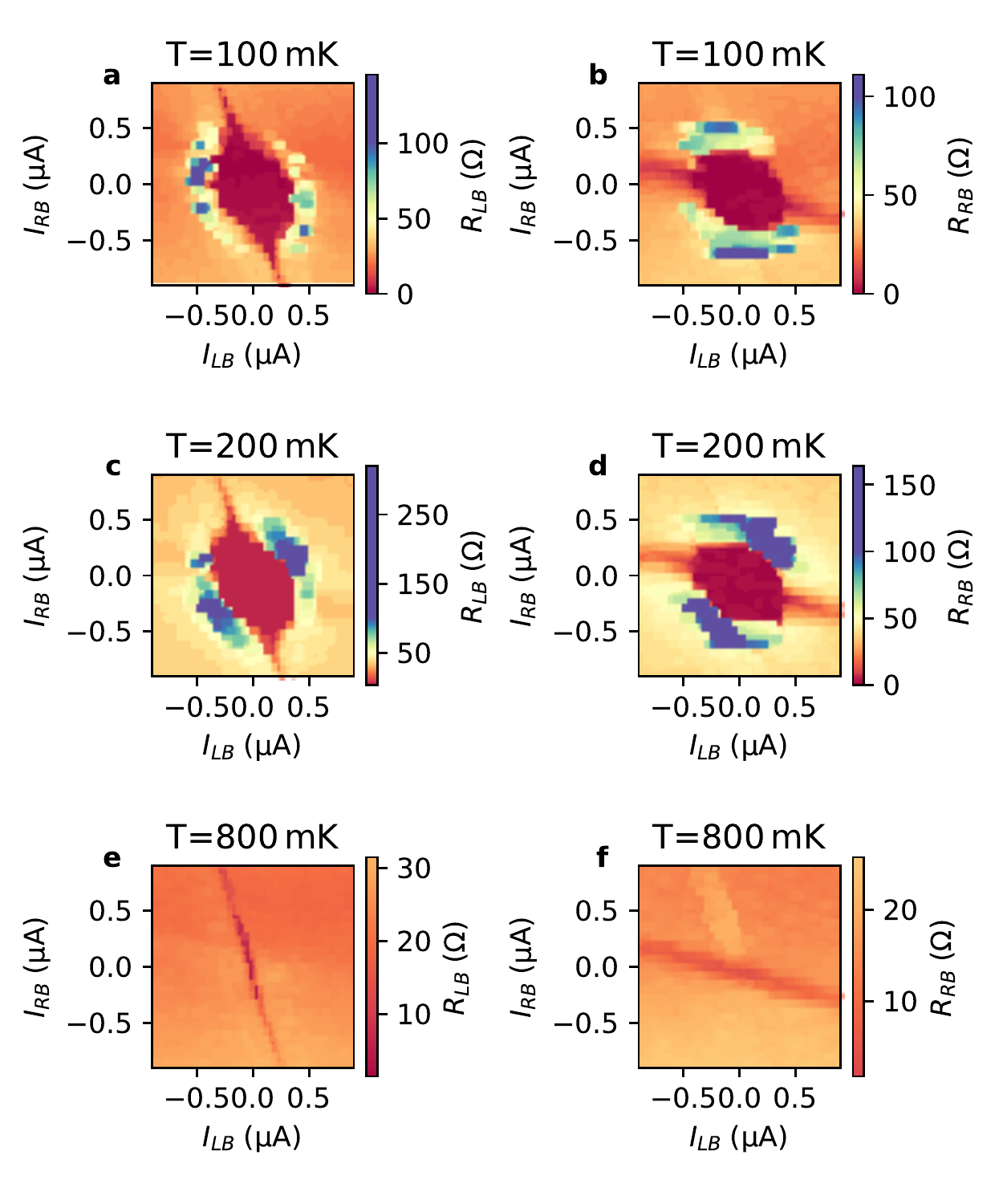}
  \caption{Differential resistance maps at various temperatures:
  Left column (a), (c), (e) shows the differential resistance $R_\mathrm{LB}$, right column (b), (d), ($\mathbf{f}$) $R_\mathrm{RB}$, accordingly. The temperatures displayed in the rows from up to down are \SI{100}{\milli\kelvin}, \SI{200}{\milli\kelvin}, and \SI{800}{\milli\kelvin}, respectively.}
  \label{fig:T_JJ_Temp}
\end{figure}

In Figs.~\ref{fig:T_JJ_Temp}(a) to (f) the differential resistance maps are shown for $R_\mathrm{LB}$ and $R_\mathrm{RB}$ measured at temperatures of 100\,mK, 200\,mK, and 800\,mK. 
One finds that with increasing temperature the area of the central superconducting region shrinks. This is in accordance with the temperature dependence of the critical current of a single Nb/Bi$_4$Te$_3$/Nb reference junction, as shown in the Supplementary Material. It is noteworthy that the superconducting feature along the inclined lines basically does not change with increasing temperature. This can be explained by the fact, that the dissipation in the neighboring junction already leads to an increased temperature being larger than the substrate temperature.

\subsection*{rf characteristics}

\begin{figure}
  \centering
  \includegraphics[width=\linewidth]{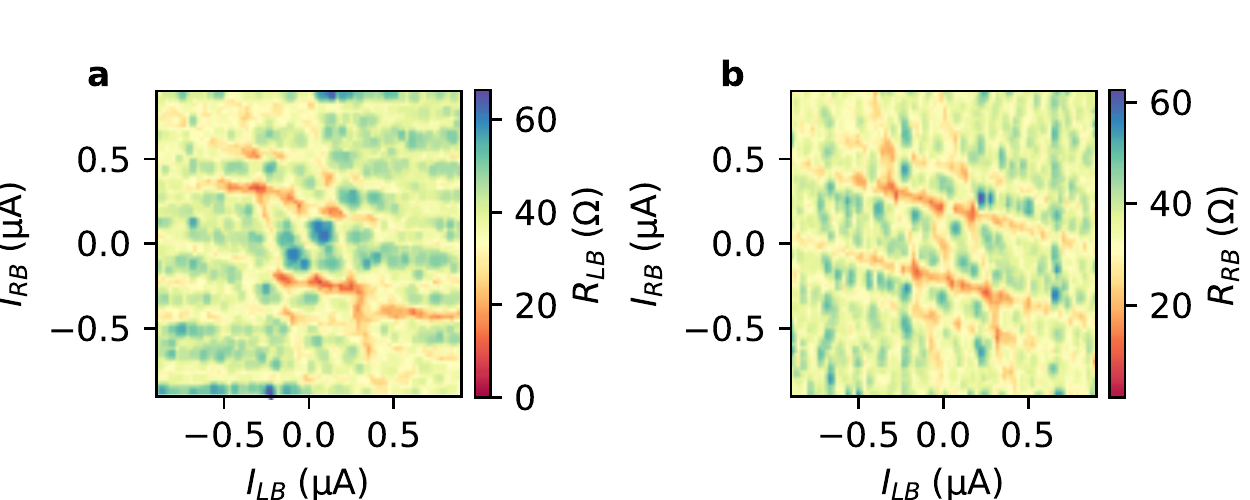}
  \caption{Shapiro step measurements at \SI{5.8}{\giga\hertz}:
  (a) Numerically determined differential resistance $R_\mathrm{LB}$ as a function of $I_\mathrm{LB}$ and $I_\mathrm{RB}$ at \SI{5.8}{\giga\hertz} and rf power of 0\,dBm. (b) Corresponding map of the differential resistance $R_\mathrm{RB}$.
  }
  \label{fig:T_JJ_5.8GHz_map_V0}
\end{figure}

Next the radio frequency response of the system is investigated in order to confirm that the experiment is described well by Josephson junction physics and to analyze the rf response of the Josephson current. This is done by first choosing a frequency and an amplitude for the rf irradiation so that both junctions show a large rf response in the differential resistance. Subsequently the same DC bias sweeps are performed as in the prior experiments. Figures~\ref{fig:T_JJ_5.8GHz_map_V0}(a) and (b) show Shapiro step measurements of the differential resistances $R_\mathrm{LB}$ and $R_\mathrm{RB}$, respectively, as a function of bias currents $I_\mathrm{LB}$ and $I_\mathrm{RB}$. The differential resistances are calculated by numerical  differentiation. Differential resistances obtained by lock-in amplifier measurements can be found in the Supplementary Material. 
The rf frequency $f_\mathrm{rf}$ and the according power was set to \SI{5.8}{\giga\hertz} and 0\,dBm, respectively. The differential resistances show clear intercrossing stripe-like patterns which can be attributed to the presence of Shapiro steps confirming the presence of a Josephson supercurrent. The intercrossing parallel stripes indicating a coupling of both junctions. By calculating the related voltage drop we find that for both junctions the Shapiro steps are located at integer multiples, $n=1,\,2, \,3 \dots$, of the characteristic voltage $V_0=h f_\mathrm{rf}/2e$. 

\begin{figure*}
  \centering
  \includegraphics[width=0.9\linewidth]{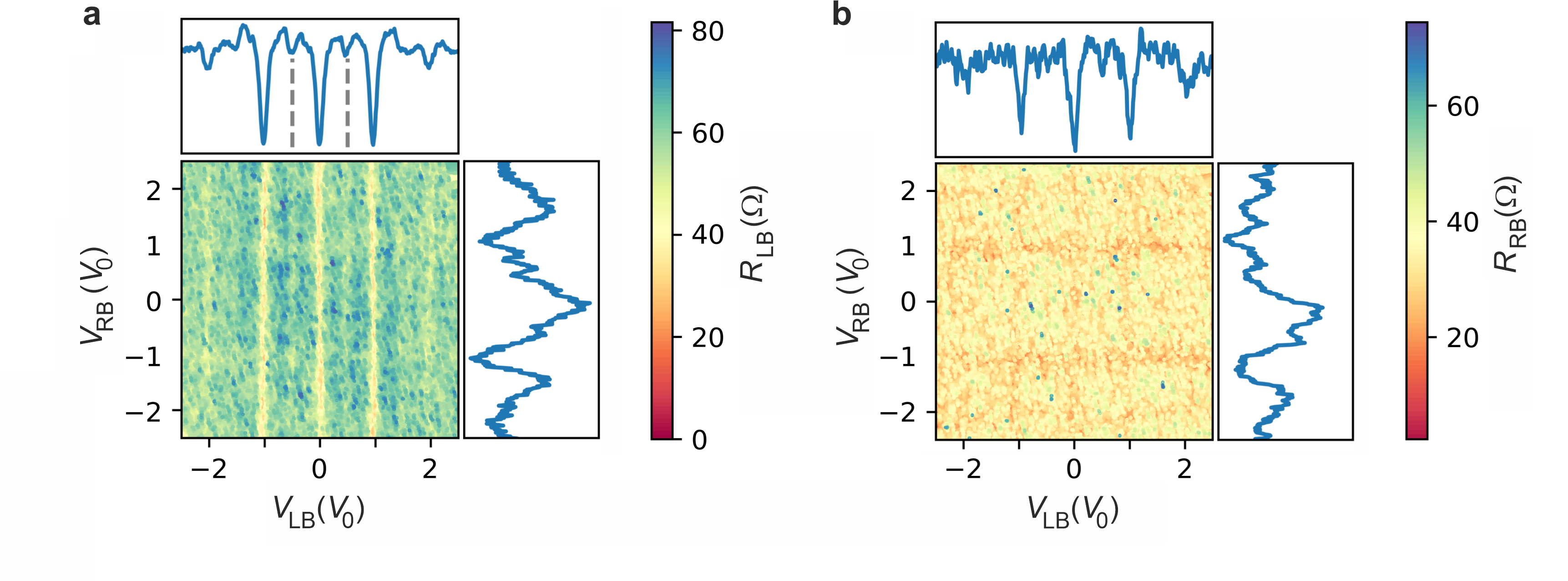}
  \caption{Shapiro step measurements at \SI{8.5}{\giga\hertz}:
  (a) Numerically determined differential resistance $R_\mathrm{LB}$ as a function of the normalized voltage drops $V_\mathrm{LB}/V_0$ and $V_\mathrm{RB}/V_0$ at \SI{8.5}{\giga\hertz} and rf power of 0\,dBm, with $V_0=hf_\mathrm{rf}/2e$. The blue curves represent the averaged signal along $V_\mathrm{LB}/V_0$ and $V_\mathrm{RB}/V_0$, respectively. The dashed lines indicate the half-integer steps. (b) Corresponding map of the differential resistance $R_\mathrm{RB}$.
  }
  \label{fig:T_JJ_8.5GHz_map_V}
\end{figure*}

In Figs.~\ref{fig:T_JJ_8.5GHz_map_V}(a) and (b) the differential resistance maps of $R_\mathrm{LB}$ and $R_\mathrm{LB}$, now taken at \SI{8.5}{\giga\hertz} at 0\,dBm, are depicted, respectively. 
Here, the color maps are plotted as a function of the normalized voltages $V_\mathrm{LB}/V_0$ and $V_\mathrm{RB}/V_0$. On first sight one finds that the Shapiro step pattern is more pronounced in $R_\mathrm{LB}$. We attribute this to a stronger coupling of the rf signal compared to the neighbouring junction due to spatial variations of the rf field. As for the measurements at \SI{5.8}{\giga\hertz} a coupling of both junctions, although weaker, is observed. Our experimental results of Shapiro step measurements are supported by comparison to simulations based on the previously introduced RCSJ model. In Supplementary Figures~4(a) and (b) maps of the simulated values of $R_\mathrm{LB}$ and $R_\mathrm{LB}$ as a function of the normalized bias voltages are shown. There, one finds that the coupling by $R_C$ results in a weak cross coupling of the Shapiro signal resulting in intercrossing stripe-like patterns of different contrast.   

A closer inspection of the resistance map presented in Fig.~\ref{fig:T_JJ_8.5GHz_map_V}(a) reveals that apart from the integer Shapiro steps also half-integer Shapiro steps, e.g. at $n=1/2$, are observed. The half integer steps are also clearly resolved in the averaged value of $R_\mathrm{LB}$ along $V_\mathrm{LB}/V_0$ shown in Fig.~\ref{fig:T_JJ_8.5GHz_map_V}(a). In single Josephson junctions such fractional steps are interpreted by assuming a skewed current-phase relationship~\cite{Snyder18,Panghotra20,Raes20} (a simulation for this case using our model is provided in the Supplementary Material). More specifically for multi-terminal junctions the rf response of superconductivity induced into normal metal has been studied previously by Duvauchelle \textit{et al.}~\cite{Duvauchelle16}. Here, half-integer steps have been found and interpreted as a feature due to the presence of coherent quartet states. However, in Fig.~\ref{fig:T_JJ_Rdiff} we did not find indications of quartet states, which would be visible by a feature in the differential resistance at opposite voltage drops on the left and right terminal \cite{Pfeffer14}. Other experimental observations of such fractional steps in multi-terminal junctions are interpreted on the basis of highly connected nonlinear networks of Josephson junctions, where (due to the higher phase space) different transitions of the phase particle in the washboard potential are possible~\cite{Arnault21}. However, since fractional Shapiro steps were observed in single junctions made with similar materials \cite{Rosenbach21a}, we favor the explanation based on a skewed current-phase relationship, which can be attributed to contributions of quasi-ballistic transport. 
In our  measurements under rf radiation we did not find indications of missing odd Shapiro steps, as predicted when Majorana bound states are present in topological junctions \cite{Dominguez17,Schueffelgen19}. Probably, for our samples the narrow width of the Bi$_4$Te$_3$ ribbons prevents the formation of these states, since due to the finite Berry phase a magnetic field along the junctions is required to gain a gap closure for the coherent surface states around the nanoribbon cross section \cite{Rosenbach21}.  

\subsection*{Magnetic field response}

\begin{figure*}
  \centering
  \includegraphics[width=1.0
 \linewidth]{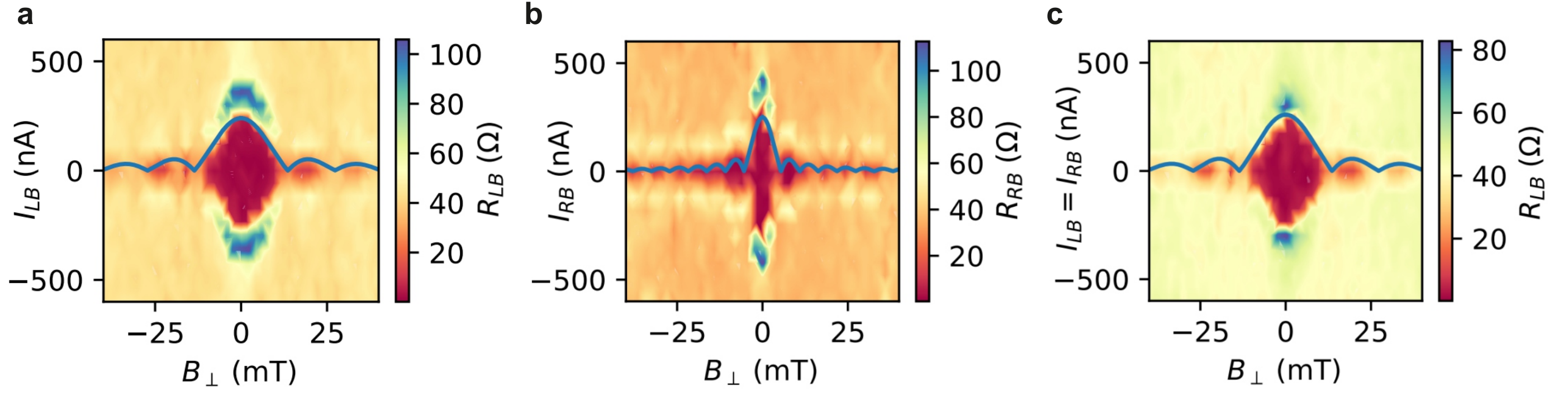}
  \caption{Differential resistances under perpendicular magnetic field sweep: (a) shows a map of $R_\mathrm{LB}$ as a function of $B_\perp$ and $I_\mathrm{LB}$ for $I_\mathrm{RB}=0$. (b) represents the corresponding map of $R_\mathrm{RB}$ as a function of $B_\perp$ and $I_\mathrm{RB}$ for $I_\mathrm{LB}=0$. In (c) $R_\mathrm{LB}$ is plotted with the sweep current chosen to be $I_\mathrm{LB}=I_\mathrm{RB}$, which corresponds a sweep along the diagonal in the current plane. In all cases a standard Fraunhofer pattern is fitted indicated as blue lines.
  }
  \label{fig:T_JJ_Fraunhofer}
\end{figure*}

The junction characteristics were also analyzed in a perpendicularly oriented magnetic field $B_\perp$.  In Fig.~\ref{fig:T_JJ_Fraunhofer}(a) the magnetic field dependence $R_\mathrm{LB}$ is plotted as a function of $B_\perp$ and $I_\mathrm{LB}$, while $I_\mathrm{RB}$ is kept at zero. One clearly observes a Fraunhofer-like interference pattern of the switching current, i.e. the boundary between the red superconducting areas and the areas with finite resistance. The blue line in Fig.~\ref{fig:T_JJ_Fraunhofer}(a) indicates the according fitting based on the Fraunhofer interference relation. The close resemblance of the experimental data to an ideal Fraunhofer pattern points towards a relatively homogeneous distribution of the supercurrent density. From the fit we extract a period of about $\Delta B=$\SI{14}{\milli\tesla}, which corresponds to a junction area of $152 \times 10^3\,\mathrm{nm}^2$. Relating these values to the dimensions of the left junction $J_\mathrm{LB}$ one finds that the period is about a factor of ten smaller than expected. Based on the actual junction size of $200 \times 72\,\mathrm{nm}^2$ a period of $\SI{144}{\milli\tesla}$ is expected for a $h/2e$ flux periodicity. We attribute the discrepancy to the experimental period to a pronounced flux focusing effect, where the magnetic field is expelled from the edge regions of the superconducting electrodes and bundled in the junction area. As a matter of fact, a comparably large flux focusing effect was previously observed in similar planar Josephson junctions based on topological insulators and Nb superconducting electrodes \cite{Rosenbach21}. 

In Fig.~\ref{fig:T_JJ_Fraunhofer}(b) the magnetic field dependence $R_\mathrm{RB}$ is shown as a function of $B_\perp$ and $I_\mathrm{RB}$ at $I_\mathrm{LB}=0$. Once again, a Fraunhofer-like interference is observed, although with a smaller period, i.e. an larger effective area where the magnetic flux is picked up. The reason for the difference compare to the measurements shown in Fig.~\ref{fig:T_JJ_Fraunhofer}(a) might be some inhomogeneity in the supercurrent density in the junction. Finally, the $R_\mathrm{LB}$ maps are scanned diagonally, i.e. $I_\mathrm{LB}= I_\mathrm{RB}$, as shown in Fig.~\ref{fig:T_JJ_Fraunhofer}(c). Here, once again a regular Fraunhofer pattern is observed, which is almost identical to the pattern shown in Fig.~\ref{fig:T_JJ_Fraunhofer}(d), indication, that the current $I_\mathrm{RB}$ through the neighboring junction basically has not effect. 

\section{Conclusion} \label{sec:discussion}

We have succeeded in extending the previously developed in situ fabrication technology for Josephson junctions to a working more complex design of a three-terminal junction. Analysis of the transport experiments shows that our system indeed behaves like a coupled network of Josephson junctions in DC transport, rf response, as well as magnetic field response. This is the first report on the topological multi-terminal devices where an interaction between the individual Josephson junctions is observed. Moreover, the observation of fractional steps in the rf response opens a window that provides a first insight into the novel physics of this type of device. On a more technical level, our results demonstrate the realization of more complex devices required for network structures in topological quantum circuits.

Further investigations and detailed understanding of such a system are crucial for the realization of complex topological quantum systems. In future, similar experiments with more intricate circuit designs and superconducting phase controlled measurements will be performed. The complexities in the junction characteristics arose from the selected weak-link material Bi$_4$Te$_3$. In future experiments we plan to incorporate conventional three-dimensional topological insulators, e.g. Bi$_2$Te$_3$, Sb$_2$Te$_3$, Bi$_2$Se$_3$, and the topological Dirac semimetal exhibited by the correctly tuned Bi$_x$Te$_y$ stoichiometric alloy.  

\section*{Acknowledgement}

We thank Herbert Kertz for technical assistance as well as Kristof Moors and Roman Riwar for fruitful discussion. This work was partly funded by the Deutsche Forschungsgemeinschaft (DFG, German Research Foundation) under Germany’s Excellence Strategy—Cluster of Excellence Matter and Light for Quantum Computing (ML4Q) EXC 2004/1—390534769, the German Federal Ministry of Education and Research (BMBF) via the Quantum Futur project “MajoranaChips” (Grant No. 13N15264) within the funding program Photonic Research Germany, as well as the Bavarian Ministry of Economic Affairs, Regional Development and Energy within Bavaria’s High-Tech Agenda Project “Bausteine für das Quantencomputing auf Basis topologischer Materialien mit experimentellen und theoretischen Ansätzen“ (grant allocation no. 07 02/686 58/1/21 1/22 2/23).\\

\clearpage
\widetext

\setcounter{section}{0}
\setcounter{equation}{0}
\setcounter{figure}{0}
\setcounter{table}{0}
\setcounter{page}{1}
\makeatletter
\renewcommand{\thesection}{S\Roman{section}}
\renewcommand{\thesubsection}{\Alph{subsection}}
\renewcommand{\theequation}{S\arabic{equation}}
\renewcommand{\thefigure}{S\arabic{figure}}
\renewcommand{\figurename}{Supplementary Figure}
\renewcommand{\bibnumfmt}[1]{[S#1]}
\renewcommand{\citenumfont}[1]{S#1}

\begin{center}
\textbf{Supplementary Material: Supercurrent in Bi$_4$Te$_3$ Topological Material-Based Three-Terminal Junctions}
\end{center}

\section{Single junction measurements}

As a reference a single Nb/Bi$_4$Te$_3$/Nb junction was measured. The junction has a length of \SI{140}{\nano\meter} and a width of \SI{500}{\nano\meter}. In Supplementary Figure~\ref{fig:t_hyst_JJ_dev}(a) the current voltage characteristics is shown at temperatures in the range from \SI{30}{m\kelvin} to \SI{0.77}{\kelvin}. At lowest temperature a critical current of \SI{750}{\nano\ampere} is obtained. In contrast to the three terminal junction, here, a hysteretic behaviour is observed, which can be explained by the missing shunt for the single Josephson junction. We attribute the hysteresis to heating resulting in a lower return current $I_r$ compared to $I_c$ \cite{Courtois08}. The critical current monotonously decreases with temperature with some kink around \SI{0.4}{\kelvin}. The latter might be attributed to a switching from a more diffusive to a more ballistic transport in the weak link \cite{Schueffelgen19}.
\begin{figure}[h!]
  \centering
  \includegraphics[width=0.98\textwidth]{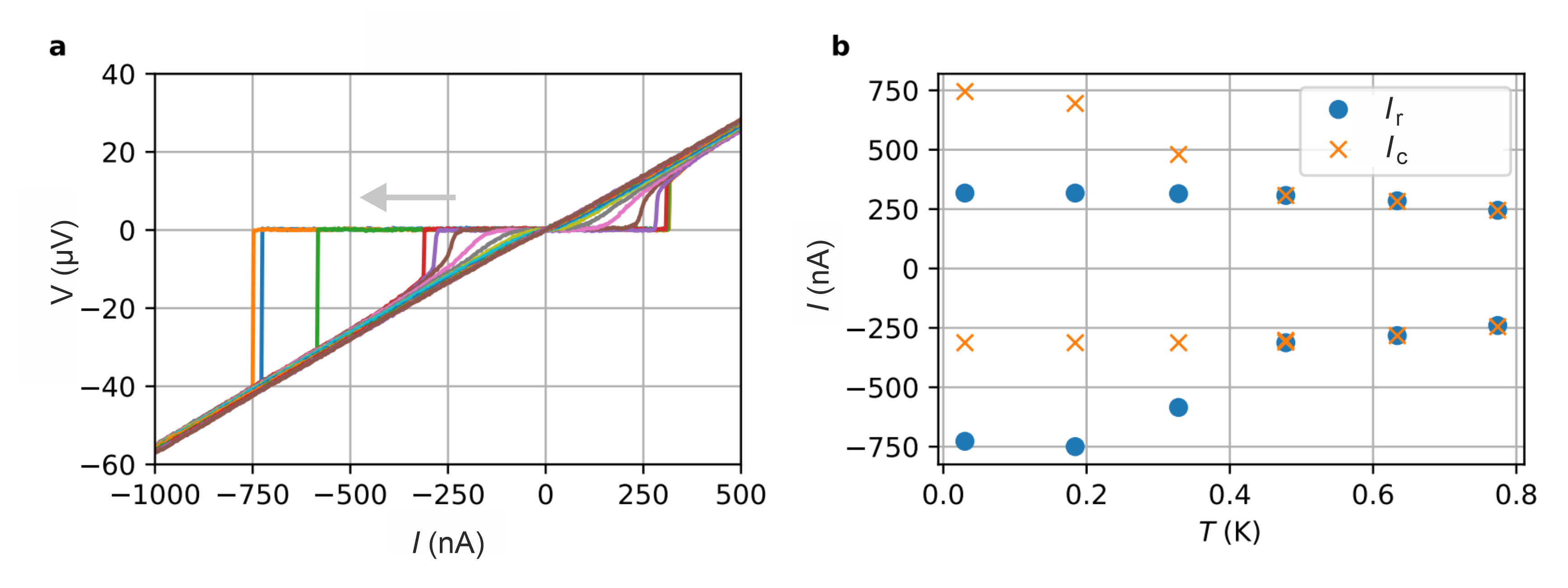}
  \caption{Current-voltage characteristics of a single Nb/Bi$_4$Te$_3$/Nb junction: (a) Current-voltage characteristics at temperatures ranging from \SI{30}{m\kelvin} to \SI{0.77}{\kelvin}. (b) Critical current $I_c$ as well as return current $I_r$ as a function of temperature.}
  \label{fig:t_hyst_JJ_dev}
\end{figure}

\section{RCSJ model for a three-terminal junction} \label{sec:RCSJ-model}  

The characteristics of our three-terminal junctions is simulated by employing a 
two-dimensional resistively and capacitively shunted Josephson junction (RCSJ) Ansatz in analogy to what was presented in previous works \cite{Graziano20,Arnault21}. In Fig.~3(a) in the main text the corresponding network is depicted including two resistively and capacitively shunted Josephson junctions with the normal state resistance $R_N$ and the capacitance $C$. We assume two identical junctions each having a critical current of $I_c$. The junctions are connected by a coupling resistor $R_C$ representing the non-superconducting junction between electrodes L and R. Following the RCSJ Ansatz the characteristics of the three-terminal junction can be described by a set of coupled differential equations of the form:
\begin{equation} \label{eq:syst1}
    \frac{I_\mathrm{LB}}{I_c} = \sin(\varphi_\mathrm{LB}) + \frac{d\varphi_\mathrm{LB}}{d\widetilde{\tau}} + \beta_c \frac{d^2\varphi_\mathrm{LB}}{d\widetilde{\tau}^2}+\frac{R_N}{R_C}\left(\frac{d\varphi_\mathrm{RB}}{d\widetilde{\tau}}-\frac{d\varphi_\mathrm{LB}}{d\widetilde{\tau}}\right) \, ,
\end{equation}
\begin{equation} \label{eq:syst2}
    \frac{I_\mathrm{RB}}{I_c} = \sin(\varphi_\mathrm{RB}) + \frac{d\varphi_\mathrm{RB}}{d\widetilde{\tau}} + \beta_c \frac{d^2\varphi_\mathrm{RB}}{d\widetilde{\tau}^2}-\frac{R_N}{R_C}\left(\frac{d\varphi_\mathrm{RB}}{d\widetilde{\tau}}-\frac{d\varphi_\mathrm{LB}}{d\widetilde{\tau}}\right) \, , 
\end{equation}
with $\varphi_\mathrm{LB}$ and $\varphi_\mathrm{RB}$ the phase differences between junctions $J_\mathrm{LB}$ and $J_\mathrm{RB}$, respectively, $ \widetilde{\tau} = t/\tau_J $ the normalized time, $\tau_J=\Phi_0/(2\pi I_c R_N)$, with $\Phi_0=h/2e $ the magnetic flux quantum, and $\beta_c= (2e/\hbar) I_c R_N^2 C$ the Stewart-McCumber parameter \cite{Tinkham04}. The equations are similar to the standard RCSJ model for a single junction, except of the last term, which introduces the current through the resistor, coupling the two junctions. This current is a result of the voltage difference between the two junctions and the coupling resistance. For $R_C\to \infty$ the coupling term goes to zero, leading to two individual junctions (decoupled system) and for $R_C\to 0$ the system is dominated by the coupling term. 

\section{Shapiro Steps in Three-Terminal Junction Experiments} 

The differential resistances $R_\mathrm{LB}$ and $R_\mathrm{RB}$ exposed to an rf radiation with a frequency of \SI{5.8}{\giga\hertz} at 0\,dBm recorded as a function of the applied DC currents are presented in Supplementary Figures~\ref{fig:R_diff_steps}(a) and (b). In contrast to the corresponding figure, which was gained by numerical differentiation, here, the resistance is directly taken using a lock-in amplifier. In Supplementary Figures~\ref{fig:R_diff_steps_8p5} the corresponding measurements at a frequency of \SI{5.8}{\giga\hertz} at 0\,dBm are shown. 
\begin{figure}[h!]
  \centering
  \includegraphics[width=0.8\linewidth]{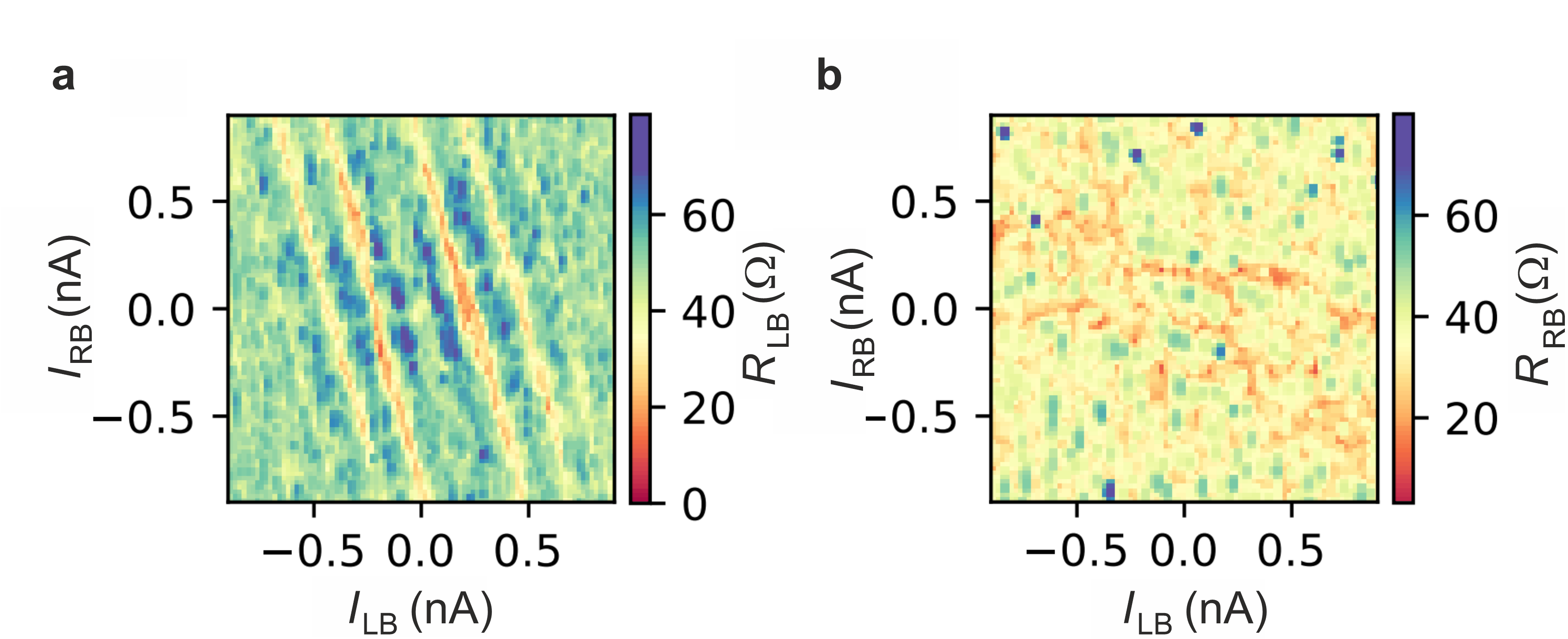}
  \caption{Shapiro Step response at 5.8\,GHz:
  (a) shows the measured differential resistance across the first junction $R_\mathrm{LB}$ as a function of the direct current $I_\mathrm{LB}$ and $I_\mathrm{RB}$ across the junction. (b) shows $R_\mathrm{RB}$ for the same current constellation.}
  \label{fig:R_diff_steps}
\end{figure}

\begin{figure}[h!]
  \centering
  \includegraphics[width=0.8\linewidth]{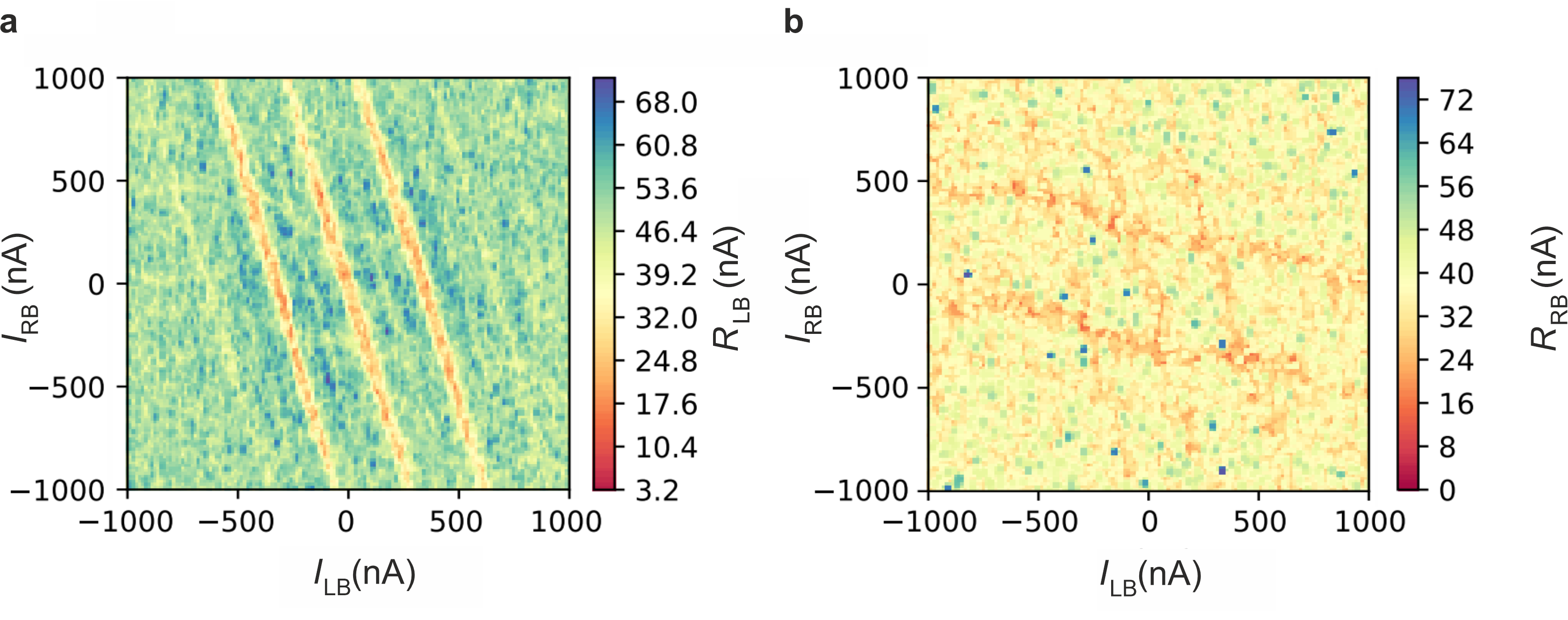}
  \caption{Shapiro Step response at 8.5\,GHz:
  $\mathbf{a}$ shows the measured differential resistance across the first junction $R_\mathrm{LB}$ as a function of the direct current $I_\mathrm{LB}$ and $I_\mathrm{RB}$ across the junction. $\mathbf{b}$ shows $R_\mathrm{RB}$ for the same current constellation.}
  \label{fig:R_diff_steps_8p5}
\end{figure}

\section{Shapiro Steps in Three-Terminal Junction Simulation} 

Using the model described in Supplementary Information~\ref{sec:RCSJ-model} the Shapiro response was simulated by adding an oscillation contribution $i_\mathrm{j,rf} \sin(2\pi f_\mathrm{rf} t)$, $j=\mathrm{LB, RB}$, to the dc bias currents. The simulated differential resistances $R_\mathrm{LB}$ and $R_\mathrm{RB}$ as a function of the normalized voltage drops at a frequency of \SI{8.5}{\giga\hertz} are presented in Supplementary Figures~\ref{fig:T_JJ_8.5GHz_sim_map_V}(a) and (b). One finds that the Shapiro response is strong in the corresponding junctions, while the coupling from the neighboring junction is weak.     
\begin{figure*}
  \centering
  \includegraphics[width=\linewidth]{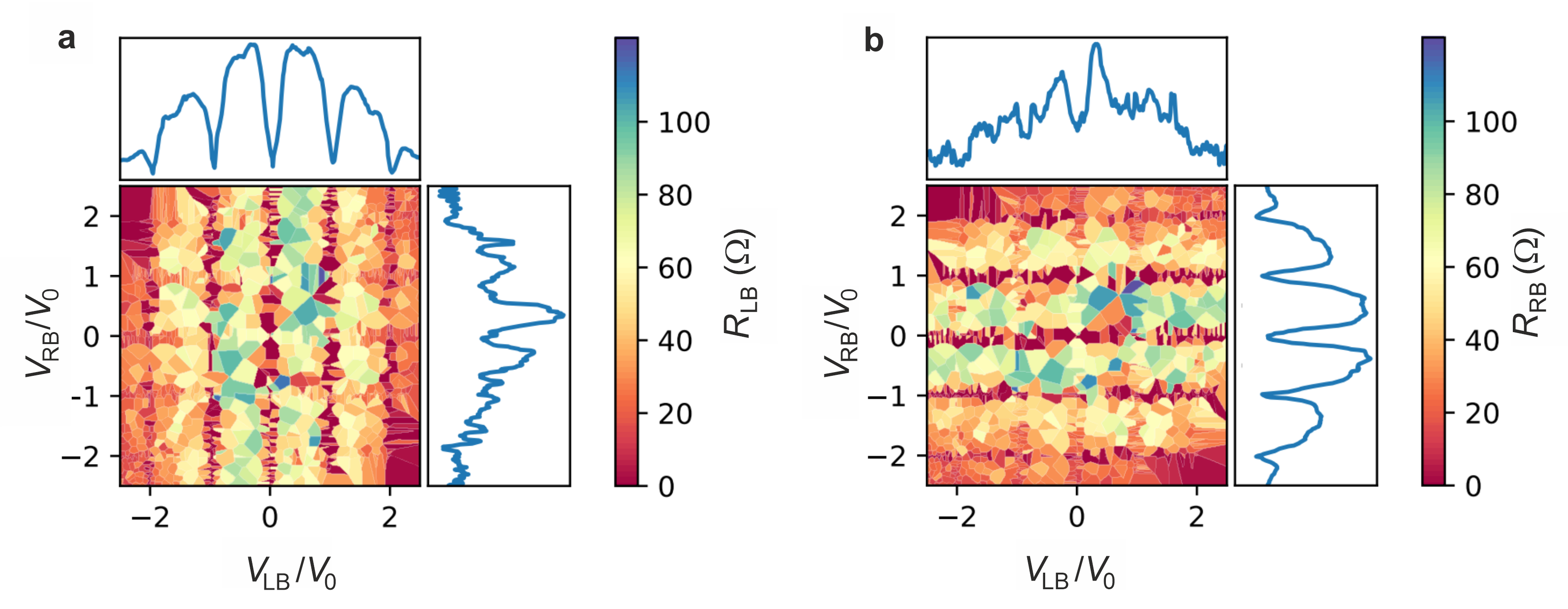}
  \caption{Shapiro step simulations at 8.5\,GHz:
  (a) Numerically determined differential resistance $R_\mathrm{LB}$ as a function of the normalized voltage drops $V_\mathrm{LB}/V_0$ and $V_\mathrm{RB}/V_0$ at \SI{8.5}{\giga\hertz}. The blue curves represent the averaged signal along $V_\mathrm{LB}/V_0$ and $V_\mathrm{RB}/V_0$, respectively. (b) Corresponding map of the differential resistance $R_\mathrm{RB}$ with the blue curves representing the averaged differential resistance along $V_\mathrm{LB}/V_0$ and $V_\mathrm{RB}/V_0$, respectively.
  }
  \label{fig:T_JJ_8.5GHz_sim_map_V}
\end{figure*}

In order to simulate the appearance of the fractional Shapiro steps a non-sinusoidal current-phase relationship was assumed for the Josephson junction by including a $\sin (2\varphi)$ contribution. In Supplementary Figures~\ref{fig:T_JJ_frac_8.5GHz_map_sim}(a) and (b) the respective simulation outcomes $R_\mathrm{LB}$ and $R_\mathrm{RB}$ for junctions $J_\mathrm{LB}$ and $J_\mathrm{RB}$ are shown as a function of bias currents. One finds that by increasing the $\sin 2\varphi$ contribution fractional steps appear. 
\begin{figure}[h!]
  \centering
  \includegraphics[width=\linewidth]{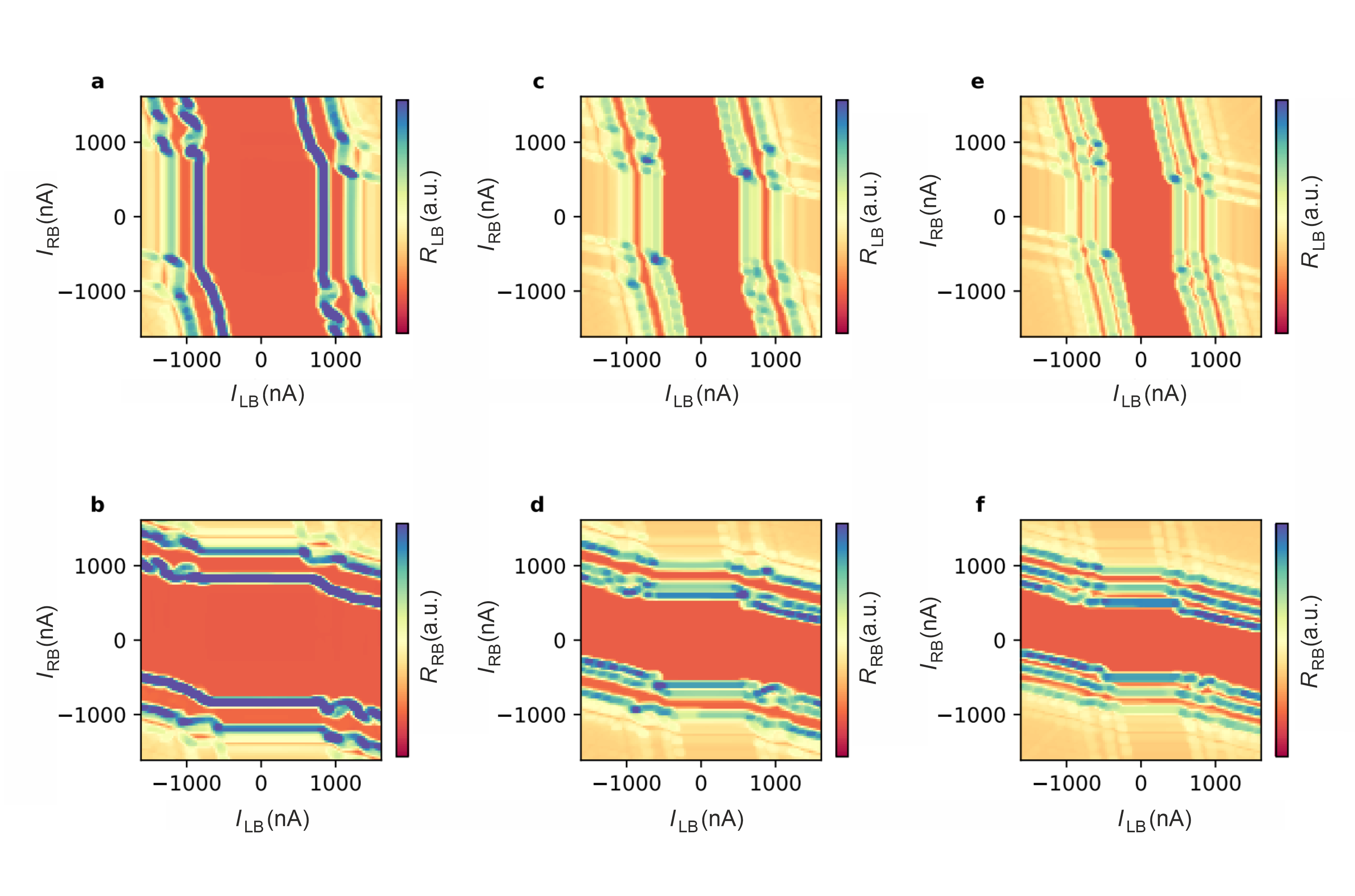}
  \caption{Simulation of Shapiro maps at \SI{8.5}{\giga\hertz} with $2\phi$ term: 
  (a) differential resistance of the first junction Shapiro steps as a function of $I_\mathrm{LB},I_\mathrm{RB}$ with an equal contribution of a $\sin{2\phi}$-term in the system, (b) the same for the second junction. (c) and  (d) show the same after doubling the $\sin{2\phi}$ contribution in the system and (e) and (f) show the same after doubling the contribution of (c) and  (d).  
  }
  \label{fig:T_JJ_frac_8.5GHz_map_sim}
\end{figure}

\end{document}